\documentstyle[prd,aps,epsfig,preprint,floats]{revtex} 
 \draft
 \tighten

\def\bort#1{}
\def\nn{\nonumber\\}
\def\eqs#1#2{\mbox{Eqs.~(\ref{#1}) and (\ref{#2})}}
\def\eq#1{\mbox{Eq.~(\ref{#1})}}
\def\be{\begin{equation}}
\def\bea{\begin{eqnarray}}
\def\eea{\end{eqnarray}}
\def\ee{\end{equation}}
\def\inv#1{\frac{1}{#1}}
\def\tr{{\rm tr\,}}
\def\sign{{\rm sign}}

\def\goto{\rightarrow}
\def\vek#1{\hbox{\boldmath$#1$}}

\def\<{\langle}
\def\>{\rangle}
\def\id{\leavevmode\hbox{\small1\kern-3.3pt\normalsize1}}
\def\simleq{\; \raise0.3ex\hbox{$<$\kern-0.75em
      \raise-1.1ex\hbox{$\sim$}}\; }
\def\simgeq{\; \raise0.3ex\hbox{$>$\kern-0.75em
      \raise-1.1ex\hbox{$\sim$}}\; }
\def\g{\gamma}
\def\ve{\varepsilon}
\def\cO{{\cal O}}
\def\apj#1#2#3{{\it Astrophys.\ J.\ }{{\bf #1} {(#2)} {#3}}}
\def\app#1#2#3{{\it Astropart.\ Phys.\ }{{\bf #1} {(#2)} {#3}}}
\def\np#1#2#3{{\it  Nucl.\ Phys.\ }{{\bf #1} {(#2)} {#3}}}
\def\pr#1#2#3{{\it Phys.\ Rev.\ }{{\bf #1} {(#2)} {#3}}}
\def\pl#1#2#3{{\it  Phys.\ Lett.\ }{{\bf #1} {(#2)} {#3}}}
\def\prl#1#2#3{{\it Phys.\ Rev.\ Lett.\ }{{\bf #1} {(#2)} {#3}}}
\def\Psibar{\overline{\Psi}}
\def\sz{\sigma_z}
\def\aell{{\bar{\ell}}}
\def\half{{\textstyle{1\over2}}}
\def\bg{\hbox{\boldmath$\gamma$}}
\def\bs{\hbox{\boldmath$\sigma$}}
\def\bB{\hbox{\boldmath$B$}}
\def\bV{\hbox{\boldmath$V$}}
\def\bP{\hbox{\boldmath$P$}}
\def\Bhat{\hbox{\boldmath$\hat B$}}
\def\bk{\hbox{\boldmath$k$}}
\def\bp{\hbox{\boldmath$p$}}
\def\bx{\hbox{\boldmath$x$}}
\def\by{\hbox{\boldmath$y$}}

\begin{document}
\thispagestyle{empty}
\title{%
\vskip0.9cm
Neutrino Dispersion in Magnetized Media and \\[1mm]
Spin Oscillations in the Early Universe}
\author{Per Elmfors}
\address{Theory Division, CERN, CH-1211 Geneva 23, Switzerland}
\author{Dario Grasso}
\address{Department of Theoretical Physics, Uppsala University,\\
Box 803, S-751 08 Uppsala, Sweden,\\
and Department of Physics, University of Stockholm,\\
Vanadisv\"agen 9, S-113 46 Stockholm, Sweden}
\author{Georg Raffelt}
\address{Max-Planck-Institut f\"ur Physik, F\"ohringer Ring 6,\\
D-80805 Munich, Germany}
\maketitle
\begin{abstract}
\hspace{-2.5mm}
We derive general expressions for the neutrino dispersion relation in
a magnetized plasma with a wide range of temperatures, chemical
potentials, and magnetic field strengths. If the electron and proton
chemical potentials vanish, as in the early Universe, there is no
magnetization contribution to the neutrino refractive index to leading
order in the Fermi coupling constant, contrary to claims in the recent
literature.  Therefore, as long as the magnetic field satisfies
$B\simleq T^2$, the neutrino refractive index in the early Universe is
dominated by the standard ``non-local term''.  If neutrinos are Dirac
particles with magnetic moment $\mu$, then their right-handed
components are thermally populated before the nucleosynthesis epoch by
magnetically induced spin oscillations if $\mu B_0 \agt
10^{-6}\mu_{\rm B}\,{\rm gauss}$, where $\mu_{\rm B}=e/2m_e$ is the
Bohr magneton and $B_0$ is a large-scale primordial magnetic field at
$T_0\approx 1\,\rm MeV$. For a typically expected random field
distribution, even smaller values for $\mu B_0$ would suffice to
thermalize the right-handed Dirac components.
\end{abstract}
\thispagestyle{empty}

\thispagestyle{empty}

\newpage
\setcounter{page}{1}
\section{Introduction}

If neutrinos carry magnetic or electric dipole or transition moments,
they can spin-precess into other spin and/or flavour states in the
presence of external magnetic fields. For example, if neutrinos were
Dirac particles with a magnetic dipole moment $\mu$, the active
left-handed states could spin-precess into the otherwise sterile
right-handed ones. It has been speculated that this effect can explain
the deficiency of the measured solar neutrino fluxes, and it certainly
can be important for supernova physics where large magnetic fields are
known to exist \cite{Raffelt}.  Further, it has been recognized for a
long time that primordial magnetic fields of sufficient strength would
couple right-handed Dirac neutrinos to the cosmic thermal heat bath
and thus cause these ``wrong-helicity'' states to be thermally
populated \cite{Lynn}. This effect would enhance the expansion rate of
the Universe at the epoch of nucleosynthesis and thus modify the
standard scenario of the formation of the light elements, in potential
disagreement with the observationally inferred abundances.

The original discussions of this cosmological effect \cite{Lynn} did
not take into account neutrino dispersion, which at that time had
received only marginal attention. Later on, it became clear that even
though the neutrino dispersion relations in vacuum and in media are
very close to that of massless particles, any deviation from the
latter may cause significant modifications of spin or 
flavour-oscillation processes.  A first assessment of medium-induced
dispersion effects for early-Universe magnetic spin oscillations was
provided in Ref.~\cite{Fukugita}. In addition, however, one has to
worry about neutrino collisions during the oscillation process. A
formalism for the simultaneous treatment of oscillations and
collisions was pioneered in Refs.~\cite{Dolgov,Stodolsky}, and was
refined in terms of quantum-kinetic equations in
Refs.~\cite{Rudzsky}. A quantum-kinetic treatment of the
early-Universe magnetic oscillation problem was provided in a recent
series of papers \cite{EnqvistS93,SemikozValle,Semikozetal,Enqvist}.

Because even fine points of the neutrino dispersion relation are
important for oscillation phenomena, one naturally wonders if the
assumed presence of a strong magnetic field may cause a spin
polarization of the electrons and positron in the medium, which in turn
may act as a new contribution to the dispersion relation. Semikoz and
Valle \cite{SemikozValle} claim that this is the case even for zero
chemical potential, and that this effect dominates the neutrino
dispersion relation for the physical conditions relevant in the early
Universe. 

Upon closer inspection, however, we find that this dispersion relation
is based on an unfortunate sign error. In a charge-symmetric plasma,
the magnetization part of the local self-energy terms cancels
between electrons and positrons rather than adding, as claimed by
Semikoz and Valle \cite{SemikozValle}.  While the correct sign can be
understood by a simple physical argument (Sect.~\ref{sss:physical-local})
and from the requirement of CPT invariance (Sect.~\ref{sss:CPT}),
  we take this opportunity to
provide the neutrino dispersion relation in a magnetized medium for
arbitrary electron chemical potential and magnetic field strength. The
correct sign is then a consequence of our completely general and
formal derivation, which leaves no room for ambiguities. Our general
expressions may also be of interest in the context of neutrino spin
oscillations in supernovae, where strong fields and very degenerate
electrons occur. Surprisingly, we find that even for arbitrary field
strengths our expressions are very similar to those derived by
D'Olivo, Nieves, and Pal \cite{DOlivoNP89} in the weak-field limit.

Neutrino dispersion in a magnetized medium may be viewed from a
somewhat different perspective where one considers an effective
neutrino electromagnetic form factor, or vertex function, induced by
the presence of the medium 
\cite{DOlivoNP89,AltherrSalati,Semikoz87,Multipoles}. Various
components of this vertex function, which is a Lorentz tensor, may be
interpreted as certain effective neutrino electromagnetic multipole
moments. 
In this language, neutrino dispersion in a magnetized medium
is represented by a medium-induced effective neutrino magnetic dipole
moment, which naturally leads to an energy shift in the presence of a
magnetic field.%
\footnote{The use of an ``effective magnetic dipole
moment'' to describe the neutrino energy shift in a magnetized medium
is somewhat misleading, because the $\g$-structure of the vertex 
function is not that of a magnetic dipole interaction. Among other
differences, only left-handed states experience any shift at all.
}   
The results of Refs.~\cite{DOlivoNP89,AltherrSalati} 
imply that in a charge-symmetric plasma this dipole moment vanishes,
in agreement with our present calculations and arguments. The same 
conclusion was reached in an early paper by Semikoz \cite{Semikoz87}, 
in conflict with the later finding of Semikoz and Valle
\cite{SemikozValle}.    

In Sect.~\ref{s:dr} we  derive general expressions for the
neutrino dispersion relation in a magnetized medium, and we derive the
relative sign of the magnetization effect by a direct physical
argument. In Sect.~\ref{s:spinosc} we investigate the efficiency of
primordial neutrino spin oscillations in view of the correct neutrino
dispersion relation in a magnetized plasma which, in the early
Universe, is well approximated by the dispersion relation of an
unmagnetized medium. Section \ref{s:disc} is devoted to a summary and
discussion.


\section{Neutrino Dispersion in Magnetized Media}
\label{s:dr}

\subsection{General Self-Energy Diagrams}
\label{ss:sediag}

\begin{figure}
\begin{center}
\epsfig{file=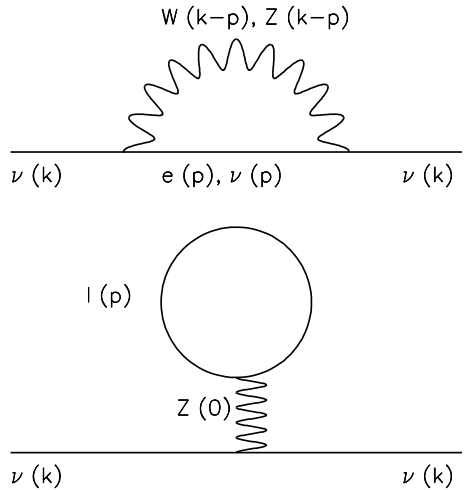,width=6cm}
\caption{The tadpole and the bubble diagrams.\label{f:diagrams}}
\end{center}
\end{figure}

In order to derive a general expression for the neutrino dispersion
relation in a magnetized medium we observe that, to lowest order, the
self-energy is given by the tadpole and bubble diagrams
shown in Fig.~\ref{f:diagrams}. To be specific we shall derive the
dispersion relation for electron neutrinos; more general cases can be
inferred by simple substitutions.

The neutrino self-energy contribution from the tadpole diagram with an
arbitrary fermion loop is
\be
\label{Sigtad}
    -i\Sigma_{\rm tadpole}=
    -\frac{1}{4}\left(\frac{ig}{\cos\theta_W}\right)^2
      \tr\!\left[\g^\alpha(c_V-c_A\g_5)iS(x,x)\right]
      \,iD^Z_{\alpha\beta}(0)\,\g^\beta L~~,
\ee
where $g$ is the weak gauge-coupling constant and $\theta_W$ the weak
mixing angle. We use the notation $R\equiv\half(1+\gamma_5)$ and
$L\equiv\half(1-\gamma_5)$. Further, $D^Z_{\alpha\beta}(\Delta)$ is
the $Z$-boson propagator while $S(x,y)$ is the coordinate-space
propagator for the background fermion. For a charged Dirac 
spin-$\inv2$
particle in the presence of an external magnetic field, $S(x,x)$ is
given in Appendix~\ref{a:loop}.  Our prime example is electrons for
which the weak coupling constants are $c_V=-\inv2+2\sin^2\theta_W$ and
$c_A=-\inv2$.
 
The bubble diagram contributes only for a background of charged
leptons from the same family as the test neutrino. In our
specific case of a test $\nu_e$ in the presence of an $e^+e^-$ plasma,
we find
\be
\label{Sigbub}
    -i\Sigma_{\rm bubble}=
    \left(\frac{ig}{\sqrt{2}}\right)^2
      R\int\frac{d^4p}{(2\pi)^4}\,\g^\alpha iS(p)\,
      iD^W_{\alpha\beta}(k-p)\,\g^\beta L~~.
\ee
For a neutrino background from the same family as the test neutrino,
there is a similar diagram with a
$Z$--$\nu$-loop that can be obtained by replacing 
$g^2\goto g^2/2\cos^2\theta_W$ and $m_W\goto m_Z$. 

The tadpole diagram provides only a local contribution, i.e.\ the
gauge-boson propagator is taken at the energy-momentum transfer
$\Delta=0$ so that we could have used an effective low-energy
four-fermion interaction.  The bubble diagram, however, involves the
gauge-boson propagator at a non-vanishing $\Delta$ so that there is a
non-local term in the self-energy.  Even in extreme astrophysical
sites, such as neutron stars, the relevant energies are so low, and
the chemical potential so high relative to the temperature, that the
bubble diagram is dominated by the local term. However, the local term
vanishes identically in a charge-symmetric plasma. Therefore, in the
early Universe the neutrino self-energy is dominated by the non-local
part of the bubble diagram \cite{NotzoldRaffelt}.


\subsection{Local Terms}
\label{ss:loc}

\subsubsection{Formal Derivation}

In order to derive the electron-neutrino dispersion relation in a
magnetized medium explicitly, we begin with the local contributions. To
this end we expand the gauge-boson propagators in powers of the
energy-momentum transfer $\Delta$,
\be
\label{gbprop}
    D^{W,Z}_{\alpha\beta}(\Delta)=\frac{g_{\alpha\beta}}{m_{W,Z}^2}+
      \frac{g_{\alpha\beta}\Delta^2-\Delta_\alpha 
      \Delta_\beta}{m_{W,Z}^4}
      +{\cal O}\left(\frac{\Delta^4}{m_{W,Z}^6}\right)~~.
\ee
The first term, which is the only one contributing to the tadpole,
gives the local part of the self-energy.  

Using the charged-fermion propagator in an external magnetic field
(see Appendix~\ref{a:loop} for more details), the tadpole
yields for a plasma consisting of electrons, neutrinos, and nucleons:
\bea
\label{tadfin}
    \Sigma_{\rm tadpole}&=&\frac{G_{\rm F}}{\sqrt{2}}\,\biggl\{
    \Bigl[-N_{n-\bar{n}}+
    2\!\!\!\sum_{i=e,\mu,\tau}\!\!\!N^L_{\nu_i-\bar\nu_i}
    -(1-4\sin^2\theta_W)(N_{e-\bar{e}}-N_{p-\bar{p}})
    \Bigr]\g_0
    \nn&&\kern21em
    +\,N^0_{e-\bar{e}}\,\Bhat\cdot\bg\,\biggr\}L~~,
\eea
where $\Bhat$ is a unit vector in the external $B$-field direction.
Further, $N_{f-\bar{f}}$ denotes the net number density of fermions
$f$, i.e.\ the total number density of fermions $f$ minus that of
antifermions $\bar{f}$. For neutrinos, only the number density of
left-handed states (superscript $L$) is counted, which is identical
to the total number density unless the right-handed degrees of
freedom have been populated by, say, magnetically induced spin
oscillations.  Usually, the standard electron and proton terms cancel
against each other in a charge-neutral plasma where
$N_{e-\bar{e}}-N_{p-\bar{p}}=0$.

In the magnetic tadpole term, $N^0_{e-\bar{e}}$ is the net number
density of electrons in the lowest Landau level.  Of course, the exact
cancellation of all higher Landau levels applies only to Dirac
fermions which do not carry anomalous magnetic dipole moments.  This
approximation is not justified for nucleons, which carry large
anomalous magnetic moments so that their polarization does not cancel
between the higher Landau levels which are not degenerate.  However,
unless the field is extremely strong or the temperature much higher
than the nucleon masses, the nucleon magnetization is suppressed by
their heavier masses relative to electrons.  Because in the present
paper we are primarily interested in early-Universe physics between
the QCD phase transition and Big-Bang nucleosynthesis (BBN), nucleons
can certainly be ignored with regard to neutrino dispersion effects.

In addition we need to consider the bubble diagram, which yields a
local contribution from electrons and electron neutrinos of 
\be
\label{locbubfin}
    \Sigma_{\rm bubble}=\frac{G_{\rm F}}{\sqrt{2}}\,2\,\Bigl[
        \left(N^L_{\nu_e-\bar{\nu}_e}
        +N_{e-\bar{e}}\right)\g_0
        -N^0_{e-\bar{e}}\,\Bhat\cdot\bg
    \Bigr]\,L~~.
\ee
Nucleons never contribute to this term.

In the weak-field limit our results agree with those found in
Ref.~\cite{DOlivoNP89}, except for the overall sign which 
is related to the convention in Ref.~\cite{DOlivoNP89} that
$e<0$ for electrons.
The approach in Ref.~\cite{DOlivoNP89}
was strictly perturbative in that a plane-wave basis for the fermions
was used instead of Landau levels.  We stress that exact expressions
for quantities such as the magnetization or the magnetic
susceptibility do not in general admit a power-series expansion in
$B$, forcing one to use Landau levels as external states
\cite{ElmforsPS95}. However, when a quantity  
does admit a power series expansion, it is not too
surprising that the linear term of the exact result agrees with a
perturbative calculation based on plane-wave states.

Our magnetic neutrino self-energy terms apply for $B\ll m_W^2$, but
$B$ may well be large compared with other scales in the problem, such as
the electron mass or the temperature.  Even for such large fields the
linear term actually gives the complete result. This surprising
finding is traced to the fact that only the lowest Landau level
contributes and that $N^0_{e-\bar{e}}$ is strictly linear in $B$.  It
must be noted, however, that the presence of the field affects the
phase-space distribution of the charged fermions and thus the
relationship between chemical potential and density. Therefore, one
must specify if the charged-particle densities or their chemical
potentials are held fixed in order to specify the functional
dependence of the neutrino dispersion relation on $B$.

The dispersion relation for left-handed electron neutrinos in a
magnetized plasma is obtained by taking the determinant of 
$\g k-\Sigma_{\rm tadpole}-\Sigma_{\rm bubble}$.
We find
\be
\label{dr}
    E_\pm=\pm k_0=\pm a+|\bk-\vek{b}|~~,
\ee
where $\pm$ refers to $\nu_e$ and $\bar{\nu}_e$, respectively. 
Further,
\bea
\label{aandb}
    \frac{a}{\sqrt{2}\,G_{\rm F}} &=&-\half N_{n-\bar{n}}+
    \sum_{i=e,\mu,\tau}\!\!\! N^L_{\nu_i-\bar{\nu}_i}
    +N^L_{\nu_e-\bar{\nu}_e} +N_{e-\bar{e}}\nn &&
    -(\half-2\sin^2\theta_W)(N_{e-\bar{e}}-N_{p-\bar{p}})~~,\nn
    \frac{\vek{b}}{\sqrt{2}\,G_{\rm F}} &=& \half
    N^0_{e-\bar{e}}\Bhat~~.  
\eea 
It is the medium- and field-induced breaking of Lorentz invariance
that generates a non-trivial dispersion relation, or refractive index,
for neutrino propagation.  In a charge-neutral plasma the term
proportional to $(N_{e-\bar{e}}-N_{p-\bar{p}})$ vanishes. Again, there
is a small nucleon contribution to $\vek{b}$ which we have
neglected. We stress that it is a slight abuse of language to call
$\vek{b}$ magnetization because only the spin part of the
magnetization enters, not the orbital part. Note further that the
spin is not a conserved quantity and only the lowest Landau level is a
spin eigenstate.


\subsubsection{Physical Derivation}
\label{sss:physical-local}

Because the local magnetization contribution to the refractive index
is controversial in the literature, it is useful to provide a more
physical derivation where the absolute sign, and the relative sign
between the electron and positron terms, become more directly
apparent. 
To this end we may start directly from the four-fermion 
neutrino vertex with a charged lepton $\ell$:
\be 
\label{Hamiltonian}
   {\cal H}_{\rm int}=\sqrt{2}\,G_{\rm F}\,
   \Psibar_\nu\g_\alpha L\Psi_\nu\,
   \Psibar_\ell\g^\alpha(g_V-g_A\g_5)\Psi_\ell~~.  
\ee 
Here, the effective weak neutral-current coupling constants $g_{V,A}$
are identical with $c_{V,A}$ unless $\ell$ is from the same family as
the neutrino, in which case $g_{V,A}=c_{V,A}+1$ because the
Fierz-transformed charged-current mimics a neutral-current
interaction. 

The neutrino self-energy is found by calculating the expectation value
$\langle \Psibar_\ell\g^\alpha(g_V-g_A\g_5)\Psi_\ell\rangle$ in a
background bath of fermions $\ell$.  In an unpolarized, isotropic
medium only the zeroth component of the vector current contributes and
yields the standard result. A magnetically induced polarization of the
charged background fermions, however, causes the axial current to
obtain a non-vanishing expectation value.

For ultrarelativistic charged fermions the expectation value
of the chirality operator $\gamma_5$ is identical with that of
$\sign(q)\hat{\bp}\cdot\Bhat\lambda$, an observation that
establishes a simple relation between chirality and the magnetic
quantum number $\lambda$ of the Landau levels.  It implies that the
axial-vector contributions cancel between charged fermions with the
same momentum but opposite $\lambda$.  The Landau-level energies 
$E^2_{n,\lambda,p_z}=m^2+p_z^2+|qB|(2n+1-\lambda)$, with
$n=0,1,2,\ldots$ and $\lambda=\pm1$, are degenerate between the levels
$(n,\lambda=+1)$ and $(n-1,\lambda=-1)$ except for the lowest level
$(n=0,\lambda=+1)$, which is not matched by a lower level with opposite
magnetic quantum number.  Therefore, only the lowest Landau level
contributes to the expectation value of the axial-vector current.

A negatively charged ultrarelativistic $\ell$ in the lowest Landau
level, moving along the $B$-field, has its magnetic moment parallel to
$\bB$, a spin opposite to $\bB$, and therefore negative
chirality. Hence for such a state
\be
  \label{Expectationvalue}
  \langle\Psibar_\ell\bg(g_V-g_A\g_5)\Psi_\ell\rangle
  =\langle \Psibar_\ell\bg\Psi_\ell\rangle(g_V+g_A)
  =\Bhat(g_V+g_A)~~.
\ee
If $\ell$ moves in the opposite direction we get $-\Bhat(g_V-g_A)$ so
that the vector part averages to zero for each momentum mode
separately if the phase-space distribution is reflection-symmetric
along $\bB$.  An anti-$\ell$ ($\aell$) moving along the $B$-field has
its magnetic moment also pointing parallel to $\bB$, but its spin in
the opposite direction relative to an $\ell$ with the same momentum
along the field, since the charge is opposite.  Therefore, the
helicity of $\aell$ is opposite to that of $\ell$ and thus their
chiralities are equal. The expectation value corresponding to
\eq{Expectationvalue} is then $-\Bhat(g_V+g_A)$.  Similarly we get
$\Bhat(g_V-g_A)$ for an $\aell$ moving in the opposite direction.
Multiplying with the net number density of $\ell$'s and $\aell$'s in
the lowest Landau level we obtain
\be
\label{Snu}
  \Sigma_\nu =-g_A\sqrt{2} G_{\rm F}
  N^0_{\ell-\bar{\ell}}\,\Bhat\cdot\bg\,L~~, 
\ee 
where the final minus sign comes from the contraction of space-like
indices in \eq{Hamiltonian}.  While our simple derivation was based on
the notion of ultrarelativistic charged leptons, this result holds
true even for non-relativistic ones as follows from the formal
derivation in the previous section.

The absolute sign of the energy shift, which differs from the one
found in Ref.~\cite{DOlivoNP89}, can be checked by comparing
\eqs{tadfin}{locbubfin} with \eq{Snu}. If the leptons are of a
family opther than the neutrinos, $g_{V,A}=c_{V,A}$, and one must
compare \eq{Snu} with the tadpole term alone. For leptons of the same family,
$g_{V,A}=c_{V,A}+1$, and the sum of the bubble and the tadpole diagram
should be compared to \eq{Snu}. The relative sign between the electron
and positron terms also follows directly from this simple derivation
without ambiguity.

In the early Universe, the numbers of particles and antiparticles are
believed to be identical to within about $10^{-9}$.  Therefore, the
plasma was effectively charge-symmetric. Our general results,
\eqs{tadfin}{locbubfin}, reveal that, to leading order in $m_W^{-2}$,
there is no magnetization contribution to the neutrino refractive
index in such an environment, contrary to what has been claimed by
Semikoz and Valle \cite{SemikozValle} who found that the fermion and
antifermion terms in \eqs{tadfin}{locbubfin} add rather than subtract.
It is correct as in Eq.~(2.9) of Ref.~\cite{SemikozValle} to identify
the relevant spatial part of the axial current
$\langle\overline\Psi_e\bg\gamma_5\Psi_e\rangle$ with the difference
between the electron and positron magnetizations, but in the
manipulations leading from their Eq.~(3.3) to (3.6) Semikoz and Valle
have unfortunately picked up an incorrect minus sign. Therefore, at
epochs before nucleosynthesis the non-local neutrino refractive terms
remain more significant than the local ones \cite{NotzoldRaffelt}, 
even
in the presence of strong magnetic fields.

\subsubsection{CPT Argument}
\label{sss:CPT}

The vanishing of the local contribution to the neutrino self-energy
in a CP-symmetric plasma can also be deduced from a direct 
symmetry argument. To this end we assume that the background plasma is
CP symmetric, and in addition we assume that it is in a stationary
state so that it is also symmetric under the time-reversal operation
T. Since CPT is strictly conserved in our theory, and the magnetic
field is CPT invariant, it follows
that neutrinos and antineutrinos of a given momentum 
must experience the same medium-induced energy shift, i.e.\ their
self-energy in the medium must be the same. Put another way, the
expectation value of $\overline\Psi_\nu \Sigma\Psi_\nu$ must be the
same for neutrino and antineutrino states of equal momenta.

At one-loop level, the general form of the self-energy operator
$\Sigma$ in a magnetized medium is \cite{DOlivoNP89} 
\be 
\Sigma = R ( a k^\mu + b u^\mu + 
c \widetilde B^\mu)\gamma_\mu L~~.
\ee 
Here, $k$ is the four-momentum of the test (anti)neutrino, $u$ is
the four-velocity of the background medium, and $\widetilde{B}_{\mu}
\equiv {1\over 2} \epsilon_{\mu\nu\alpha\beta}u^{\nu} F^{\alpha\beta}$
is a covariant expression for the external electromagnetic field which
is a pure B-field in the rest frame of the medium.  The coefficients
$a$, $b$, and $c$ are functions of the scalars $k^{2}$, $\omega \equiv
k\cdot u$, and $k\cdot \widetilde{B}$.

Under CPT the current $\overline\Psi_\nu \gamma^\mu\Psi_\nu$ and the
four-momentum $k$ change sign. (Recall that the Dirac eq. implies
$\overline\Psi_\nu k^\mu\gamma_\mu\Psi_\nu = m \overline\Psi_\nu\Psi_\nu$,
and that $\overline\Psi_\nu\Psi_\nu$ is invariant 
under CPT.) However, the four-vectors $u$ and $\widetilde{B}$ are
invariant under CPT. It is important to observe here that $u$
is not an operator for $\Psi_\nu$ since it is just fixing the
new reference frame. 
From \eqs{Sigtad}{Sigbub} it is clear that the
local contribution to $\Sigma$ is independent of $k$ so that the
coefficient $a$ must be zero while $b$ and $c$ must be
constants. However, because $u^\mu$ and $\widetilde B^\mu$ are even
under CPT while $\langle\overline\Psi_\nu\gamma_\mu\Psi_\nu\rangle$ is
odd, and because $\langle\overline\Psi_\nu\Sigma\Psi_\nu\rangle$ is
required to be even, we find that $b$ and $c$ must be zero.  The
coefficient $c$ is related to a medium-induced effective neutrino
magnetic dipole moment.  In Ref.~\cite{NievesP89} it was already shown
on the basis of the same argument that such a dipole moment must
vanish.
Thanks to this argument no contributions to $\Sigma_{\rm local}$ can
arise from strong-field corrections to the $W$ propagator
in a CPT symmetric plasma.
Non-local terms which are odd functions of scalars that are odd under
CPT, namely $\omega$ and $k\cdot \widetilde{B}$, are not required to
vanish.


\subsection{Non-local Terms}
\label{ss:nonloc}

In a charge-symmetric plasma, all of the local self-energy terms given
in \eq{aandb} vanish so that the second term in the expansion of the
gauge-boson propagator in \eq{gbprop} dominates.  We shall concentrate on
the case where $m_e\ll T\ll m_W$ and $B\simleq T^2$. In the early
Universe, these are quite reasonable approximations between the QCD
phase transition and nucleosynthesis.  Repeating the calculations for
the bubble diagram we then obtain 
\bea
\label{Signonloc}
    \Sigma_{\rm bubble}(k)
    &=&-\frac{7\sqrt{2}\,\pi^2G_{\rm F} T^4}{45\,m_Z^2}
    \left(1+\frac{2m_Z^2}{m_W^2}\right)
    \left(\g_0 k_0 -\inv4 \g k\right) L
    \nn &&
    -\frac{\sqrt{2}\,G_{\rm F} T^2}{6\,m_W^2}\,e
    \bB\cdot\bs
    \left[\g_0k_0+(\Bhat\cdot\bg)
    (\Bhat\cdot\bk)\right]\,L~~,
\eea 
where $\bs$ is a vector of Dirac spin matrices
defined by $\bB\cdot\bs=\frac{i}{4}F^{\mu\nu}[\g_\mu,\g_\nu]$,
where $F^{\mu\nu}$ is the field strength tensor.
The resulting dispersion  relation takes the form
\bea
\label{drnl}
    E_\pm=\pm k_0&=&\left[1-\frac{7\sqrt{2}\,\pi^2G_{\rm F} T^4}
    {45\,m_Z^2}
      \left(1+\frac{2\,m_Z^2}{m_W^2}\right)\right]|\bk|
    \pm\frac{\sqrt{2}\,G_{\rm F} T^2}{3\,m_W^2}\,
    e\bB\cdot\bk \nn
    &\approx&\left(1-6.0\,\frac{G_{\rm F}T^4}{m_W^2}\right)|\bk|
    \pm0.47\,\frac{G_{\rm F} T^2}{m_W^2}\,e\bB\cdot\bk~~.
\eea
The first part agrees with the result of Ref.~\cite{NotzoldRaffelt};
it is the same for $\nu_e$ and $\bar{\nu}_e$.  The $B$-dependent
energy shift is anisotropic and opposite for $\nu_e$ and
$\bar{\nu}_e$. However, it remains subdominant compared to the
isotropic term as long as $B\simleq T^2$.
 

\section{Spin Oscillation in the Early Universe}
\label{s:spinosc}

\subsection{Neutrino Depolarization Rate}

As an application of our results we consider explicitly the case of
Dirac neutrinos with a magnetic moment $\mu$. In the presence of a
primordial magnetic field the thermally populated left-handed
(l.h.)  states can spin-precess into the otherwise sterile
right-handed (r.h.)  ones, thus which will be populated as well. This
process of populating the ``wrong-helicity'' neutrino states is
treated theoretically by virtue of a Boltzmann-type kinetic equation,
which includes neutrino oscillations as discussed in
Refs.~\cite{Rudzsky}. However, for a simple estimate one may ignore
the detailed evolution of the individual momentum modes and rather
study an average evolution of the overall spin-polarization vector
$\bP$ of the entire ensemble.  In this simplified approach the
global spin-polarization vector evolves as \cite{Stodolsky}
\be
    \label{X001}
    \partial_t\bP=\bV\times \bP-D \bP_{\rm T}~~,
\ee
where $\bV$ is a vector of effective magnetic interaction
energies, $D$ a damping rate due to collisions, and $\bP_{\rm T}$
the ``transverse'' part of the spin-polarization vector to be
discussed below.

In the absence of a medium, the damping rate vanishes and the
effective interaction energy for ultrarelativistic neutrinos is
$\bV=2\mu \bB_{\rm T}$, where $\bB_{\rm T}$ is the component of the
$\bB$ field which is transverse to the neutrino direction of motion
\cite{Raffelt,Shrock}. Because only the transverse magnetic field
matters, in vacuum the neutrino helicity can be reversed entirely by
spin precessions. Put another way, l.h.\ and r.h.\ states are
maximally mixed by the presence of a magnetic field, independently of
the field direction with respect to the neutrino direction of motion,
unless $\bB_{\rm T}$ vanishes exactly. Of course, the precession time
depends on the magnitude of $\bB_{\rm T}$ and thus on the relative
field direction.

The first impact of a medium is that it endows the active (l.h.)
neutrino states with a nontrivial dispersion relation, while the
sterile (r.h.) ones remain unaffected.  Because the
particle-antiparticle asymmetry in the early Universe is thought to be
of order $10^{-9}$ for all species, i.e. small, the dominant
contribution to the neutrino refractive index is the non-local term
that was first identified in Ref.~\cite{NotzoldRaffelt}.  The
discussion in Sect.~\ref{ss:nonloc} reveals that even in a magnetized
charge-symmetric plasma, the additional neutrino refractive term from
the $B$ field is rather small so that the standard isotropic term
continues to dominate, in agreement with the treatment
of~Ref.~\cite{Fukugita}. We expect this to remain true, and that the
first term of \eq{Signonloc} remains good as an approximation even for
temperatures not much higher than the electron mass which are relevant
at the time of the BBN. However, in Ref.~\cite{Fukugita}
the impact of the damping term was not properly discussed. The
later systematic studies of kinetic equations for oscillating
neutrinos were not available at that time. 

In order to identify $\bV$ and $D$ relevant for the conditions of the
early Universe we begin with the energy difference between l.h.\ and
r.h.\ neutrinos of flavour $\ell=e$, $\mu$, or $\tau$ which is $E_{\rm
l.h.}-E_{\rm r.h.}=-\xi E$. Here, $E$ is the unperturbed energy, which
agrees with $E_{\rm r.h.}$ because r.h.\ neutrinos do not experience
any energy shift in the medium. 
Assuming the mass of the neutrino to be much smaller than the
temperature, it is easy to extract the coefficient $\xi$
{}from \eq{drnl}: 
\be
    \label{X003}
    \xi=\frac{8\sqrt2\,G_{\rm F}}{3}
    \left(\frac{\rho^L_{\nu_\ell+\bar{\nu}_\ell}}{m_Z^2}+
      \frac{\rho_{\ell+\bar{\ell}}}{m_W^2}\right)~~.
\ee
Here $\rho^L_{\nu_\ell+\bar{\nu}_\ell}$ is the energy density in
l.h.\ neutrinos plus antineutrinos of flavour $\ell$ while
$\rho_{\ell+\bar{\ell}}$ is the energy density in the $\ell$-flavoured
charged leptons plus antileptons. The coefficient
 $\xi$ has the same sign
for neutrinos and antineutrinos as test particles.  For $\tau$
neutrinos in the early Universe it is dominated by the
$\rho^L_{\nu_\tau+\bar{\nu}_\tau}$ term because the presence of
$\tau$ leptons is suppressed by a Boltzmann factor $e^{-m_\tau/T}$.

In a magnetized charge-symmetric plasma the
spin-polarization vector of an ultrarelativistic neutrino or
antineutrino of energy $E$ evolves according to Eq.~(\ref{X001}) with
\bea
    \bV_{\rm T}&=&2\mu\bB_{\rm T}~~,\nonumber\\
    |\bV_{\rm L}|&=& \xi E~~,
\label{X004}
\eea
where T and L are understood to be transverse and longitudinal
relative to the neutrino direction of motion.  Since we
are using neutrino helicity states, the direction of spin-quantization
is identical with 
the direction of motion.  Therefore, the tilt of $\bV$
relative to the direction of motion is twice the effective in-medium
mixing angle \cite{Stodolsky} between l.h.\ and r.h.\ states:
\be
\label{X005}
    \tan 2\theta=\frac{V_{\rm T}}{V_{\rm L}}
    =\frac{2\mu B_{\rm T}}{\xi E}~~,
\ee
where $V_{\rm T,L}=|\bV_{\rm T,L}|$ and $B_{\rm T}=|\bB_{\rm T}|$.
Thus, for sufficiently weak magnetic fields the l.h.\ and r.h.\ states
are effectively de-mixed so that l.h.~states spin-precess only
partially into r.h.~ones. Put another way, the spin precession is
about the direction of an effective magnetic field $\bB_{\rm
eff}\equiv \bV/\mu$ which is no longer transverse to the direction of
motion.

The second effect of a medium is that l.h.\ neutrinos scatter, thereby
interrupting the precession process. A collision essentially amounts
to a ``measurement'' of the helicity content of a given neutrino
because the l.h.\ component is scattered out of its previous
direction of propagation while the r.h.\ component moves on
unscathed. This implies that every collision resets the neutrino into
a helicity eigenstate and the oscillation process begins from
scratch. Collisions thus destroy the phase coherence between the l.h.\
and r.h.\ component of a neutrino state, which amounts to a damping of
the transverse part $\bP_{\rm T}$ of the polarization vector. In the
present situation where the r.h.\ component does not interact at all,
the damping rate $D$ is found to be {\it half\/} the collision rate of
the l.h.\ component \cite{Stodolsky,Rudzsky} so that, in the early
Universe:
\be
    \label{X006}
    D=f_{\rm D}\,\frac{7\pi}{48}\,G_{\rm F}^2 T^4 E~~.
\ee
Here, $f_{\rm D}$ is a numerical factor, which was
found to be unity for $\mu$- or $\tau$-flavoured (anti)neutrinos in a
background medium of $e^\pm$ and all sequential (anti)neutrinos
\cite{NotzoldRaffelt}.  Corrections from the magnetic field are
expected to be small for $eB\simleq T^2$.

For an estimate of the rate of depolarization $\Gamma_{\rm depol}$ of
the initially l.h.\ (anti)neu\-trino population, we turn to thermal
averages of the refractive and damping terms. The average energy is
$\langle E\rangle\approx 3T$ for a given neutrino species where the
equality would be exact if the neutrinos would follow a
Maxwell-Boltzmann distribution instead of a Fermi-Dirac one. Then 
\be
\label{X007} 
    \langle D\rangle=f_{\rm D}\,\frac{7\pi}{16}\,G_{\rm F}^2 T^5 
\ee 
is the average damping term. For the refraction term in \eq{X003}
we note that the energy density in one flavour $\nu_\ell$ of l.h.\
(anti)neutrinos is
$\rho_{\nu_\ell+\overline\nu_\ell}=(7\pi^2/120)\,T^4$.  Further,
$m_Z^{-2}=(\sqrt2\,G_{\rm F}/\pi\alpha) \sin^2\theta_W\cos^2\theta_W$
where $\alpha\approx1/137$ is the fine-structure constant. We will
always approximate $\sin^2\theta_W=1/4$. With $\langle E\rangle=3T$ we
thus find 
\be 
\label{X008} 
    \langle V_{\rm L}\rangle\approx f_{\rm L}\,
        \frac{7\pi}{40\alpha}\,G_{\rm F}^2\,T^5~~, 
\ee 
where $f_{\rm L}\equiv1+(\rho_{\ell+\bar{\ell}}/
\rho^L_{\nu_\ell+\bar{\nu}_\ell}) \,(m_Z/m_W)^2$ is a factor to
account for the possible presence of charged leptons of flavour $\ell$,
which would also contribute to Eq.~(\ref{X003}).  For $\mu$ and $\tau$
neutrinos the lepton densities are small and we have $f_{\rm L}=1$,
while for $e$ neutrinos $f_{\rm L}\approx 3.6$.

These results are enough to determine that the evolution 
Eq.~(\ref{X001}) of the neutrino polarization vector is weakly damped,
i.e.\ that it typically precesses several 
times between collisions. The
oscillation or spin-precession frequency is identical to
$V=(\bV_{\rm T}^2+\bV_{\rm L}^2)^{1/2}>V_{\rm L}$.  
With Eqs.~(\ref{X007}) and (\ref{X008}) we find
\be
    \label{X009}
    \frac{\langle D\rangle}{\langle V_{\rm L}\rangle}
    \approx\frac{f_{\rm D}}{f_{\rm L}}\,\frac{5\alpha}{2}~~. 
\ee
The average period of spin precession is approximately
$2\pi/\langle V_{\rm L}\rangle$ so that there are at least about 10
revolutions between collisions.

In order to understand the solution of Eq.~(\ref{X001}) in the
weak-damping limit we multiply both sides with $\bP/P^2$, which
leads to $\partial_t P/P=-D\,(P_{\rm T}/P)^2$ with $P=|\bP|$ and 
$P_{\rm T}=|\bP_{\rm T}|$. In the weak-damping limit we may use a
precession-averaged $\overline P_{\rm T}$, which is found by taking the
transverse part of the projection of $\bP$ on $\bV$.
Elementary geometry yields $\overline P_{\rm T}/P=
\cos2\theta\sin2\theta$ so that in the weak-damping limit
\be
\label{X010}
    \partial_t P/P =-D\,\cos^22\theta\,\sin^22\theta~~,
\ee
where the effective mixing angle is given by Eq.~(\ref{X005}). 
Assuming that it is small, so that $\cos2\theta\approx1$
and $\sin2\theta\approx\tan2\theta$ we thus find
\be
\label{X011}
    \Gamma_{\rm depol}\approx
    \frac{(2\mu B_{\rm T})^2\,\langle D\rangle}
    {\langle V_L^2\rangle}
    \approx \frac{f_{\rm D}}{f_{\rm L}^2}\,
    \frac{400\,\alpha^2}{7\pi}\,  
    \frac{\mu^2B_{\rm T}^2}{G_{\rm F}^2 T^5}
\ee
for the average depolarization rate
$\Gamma_{\rm depol}\equiv\langle D\cos^22\theta\sin^22\theta\rangle$.
We have freely factorized the thermal averaging process, and we have
assumed a homogeneous magnetic field.


\subsection{Comparison with Expansion Rate}
 
In order to decide whether the depolarization of the initially l.h.\
neutrino ensemble is ever complete during the cosmic evolution, we need
to compare $\Gamma_{\rm depol}$ with the expansion rate $H$. If at
some epoch $\Gamma_{\rm depol}\agt H$ then r.h.\ neutrinos have
approximately reached thermal equilibrium at that time. If this epoch
falls between the QCD phase transition at $T_{\rm QCD}\approx150\,\rm
MeV$ and nucleosynthesis at $T_{\rm BBN}\approx1\,\rm MeV$, then a
significant impact on the primordial light-element abundances would
have to be expected.  

According to the Friedmann equation the expansion rate is given by
$H^2=(8\pi/3)\,\rho/m_{\rm Pl}^2$, where $\rho$ is the energy density
of the Universe at a given epoch and $m_{\rm
  Pl}=1.22\times10^{19}\,\rm GeV$ is the Planck mass.  In the
radiation-dominated epoch, the energy density is
$\rho=g\,(\pi^2/30)\,T^4$, with $g$ the effective number of thermally
excited degrees of freedom. 
Between the QCD and BBN epochs we need to
count photons, $e^\pm$, and the l.h.\ sequential (anti)neutrinos so
that $g=43/4$.  Therefore, we need to require that at some epoch
$\Gamma_{\rm depol}$ exceeds
\be
   \label{X012}
   H=f_H (43\pi^3/45)^{1/2}\,T^2/m_{\rm Pl}~~,
\ee
where $f_H\equiv (g\,4/43)^{1/2}$. 
The energy density of the magnetic field should be added to $\rho$,
but since it is at most of the same order of magnitude 
as the radiation energy it can be absorbed into $f_H$ without changing
our analysis significantly.

In order to perform this comparison we need to understand the scaling
law of an assumed magnetic field distribution under the cosmic
expansion. Flux conservation indicates that $B\propto R^{-2}$ or
$B\propto T^2$. This would be the complete scaling if the magnetic
field were homogeneous. In practice, any primordial field distribution
is expected to be complicated and noisy, so that the effective $B_{\rm
T}^2$ in \eq{X011} must be interpreted as a suitable average.
Essentially, each neutrino oscillation process ``measures'' the
magnetic field linearly averaged over distances corresponding to the
oscillation length so that we need to estimate 
$\<\bB_{\rm T}\>_{L_{\rm osc}}\equiv
\inv L_{\rm osc}\int_{L_{\rm osc}} |dx| \bB_{\rm T}(x)$, 
where the integral is over a neutrino
oscillation path. This linear average is relevant because the
spin-precession equation (\eq{X001}) is linear in $\bB_{\rm T}$.
Further, one must take an ensemble average over all oscillation paths
at a given epoch.  Here, the average should be taken over the quantity
$\langle \bB_{\rm T}\rangle_{L_{\rm osc}}^2$ since the expression for
the depolarization rate in \eq{X011} is quadratic in $B_{\rm T}$.  The
co-moving oscillation length $L_{\rm osc}$ increases with time so that
at later times the effective field strength is averaged over larger
co-moving domains, reducing the effective field strength below the
naive $T^2$ scaling law.

As a simple model for the $B$-field scaling we assume a power law 
in co-moving coordinates of the form \cite{Enqvist}
\be
\label{Bscaling}
    \<\!\<B_{\rm T}\>\!\>_L\equiv \<\,\< \bB_{\rm T}\>_L^2\>^{1/2}
    =B_d\left(\frac{T}{T_0}\right)^{2}
    \left(\frac{d}{L}\right)^\gamma~~, 
\ee 
where $T_0$ is the temperature at a reference epoch which we take to
be the BBN time with $T_0=1\,{\rm MeV}$, $d$ is the co-moving size of
a typical domain over which the field is correlated, and $B_d$ is the
field strength in such a domain. The average $\<\!\<\ldots\>\!\>_L$
indicates a linear averaging over the length scale $L$ and a 
root-mean-square
average over all oscillation paths.  The exact
$B$-field scaling depends on the mechanism of initial generation and
the evolution of the complicated magnetohydrodynamic equations
\cite{BEO}. Therefore, the power law in \eq{Bscaling} should only be
taken as a toy model for $L\gg d$. In particular, the merging of two
domains would change both $d$ and $B_d$ while we take the combination
$B_d T_0^{-2} d^\gamma$ to be constant here.

Because oscillating neutrinos measure the magnetic field over an
oscillation length scale, it is natural to use $L=L_{\rm osc}$, which 
is of order $H^{-1}$ at BBN . Therefore, we define
a horizon-scale magnetic field at BBN by 
$B_0\equiv\<\!\< B_{\rm T}\>\!\>_{L_{\rm osc}}$ taken at $T=T_0$. 
The condition $\Gamma_{\rm depol}\agt H$ for which r.h.\
neutrinos are certain to reach thermal equilibrium at some epoch $T$
then translates into
\be
\label{X013}
   \mu B_0\agt\left(\frac{f_H f_{\rm L}^2}{f_{\rm D}}\right)^{1/2}\,
   \frac{\pi}{20}\,\left(\frac{2107\,\pi}{45}\right)^{1/4}
   \frac{G_{\rm F} T^{3/2}\,T_0^{2}}
   {\alpha\,m_{\rm Pl}^{1/2}}
   \left(\frac{L_{\rm osc}(T)}{L_{\rm osc}(T_0)}\right)^\gamma~~.
\ee
The temperature dependence of the co-moving oscillation length can be 
determined
from $|{\bf V}|^{-1}\simleq V_{\rm L}^{-1}$ in \eq{X008} so that
$L_{\rm osc}(T)/L_{\rm osc}(T_0)=(T_0/T)^4$.
If we focus on the
period between the QCD phase transition and BBN we have $f_H=1$, and
the numerical coefficient in \eq{X013} is $0.55\approx 1/2$. Then
\be
\label{bound}
   \mu B_0\agt\frac{G_{\rm F} }
   {2\alpha\,m_{\rm Pl}^{1/2}}\,
   T^{3/2-4\gamma}\,T_0^{2+4\gamma}~~,
\ee
where in addition we have used $f_{\rm D}=f_{\rm L}=1$ appropriate 
for $\nu_\mu$ and $\nu_\tau$.

Evidently we need to distinguish between two generic cases depending
on whether $\gamma<3/8$ or $\gamma>3/8$. Beginning with the former,
which would be applicable for a homogeneous field,
the condition in \eq{bound} should be imposed at as low a temperature
as possible in order to find the smallest necessary $\mu B_0$ 
sufficient to populate the r.h.\ states.  Our entire discussion makes
sense only as long as neutrinos scatter efficiently by weak
interactions. At later times they may still spin-precess in the cosmic
magnetic field, but this would have no further impact on the expansion
rate of the Universe as the only effect would be to redistribute the
total neutrino energy density between r.h.\ and l.h.\ states.
Neutrinos freeze out at about $T=1\,\rm MeV$, just before the BBN
epoch. With
$T=T_0=1\,\rm MeV$ r.h.\ neutrinos reach thermal equilibrium before
BBN if
\be
   \label{X014}
   \mu B_0 \agt 
   \frac{G_{\rm F} T_0^{7/2}}{2\alpha\,m_{\rm Pl}^{1/2}}
   \approx 7\times 10^{-15}\,{\rm eV}\approx
   1.2\times10^{-6}\mu_{\rm B}\,{\rm gauss}~~.
\ee

This requirement is essentially identical to what was found in
Ref.~\cite{Fukugita}, even though the interplay between oscillations
and collisions was not treated there.  It was demanded that the mixing
angle should be large, and that the damping rate should be small
compared with the spin-precession rate, conditions which are
sufficient, but not necessary to achieve thermal equilibrium. Here we
found that we are always in the weak-damping case. If we take damping
effects into account according to Eq.~(\ref{X001}), the required
magnitude for $\mu B_0$ at the critical epoch around neutrino 
freeze-out implies that the mixing angle is not small.  Therefore, either
treatment leads to roughly the same answer.  The underlying reason for
this coincidence is that in the early Universe the dispersive and the
absorptive parts of the neutrino refractive index are of the same
general magnitude, i.e.\ they are both second order in $G_{\rm F}$.
Then, at the critical epoch around neutrino freeze-out, the time
scales $\langle V_{\rm L}\rangle$, $\langle D\rangle$, and $H$ are all
about the same to within numerical factors.
 
The assumption that the magnetic field is a slowly varying function
($\gamma\approx 0$) on the scale of the Hubble radius at
nucleosynthesis is not physically very likely. In fact, an ubiquitous
{\it mean field\/} would be incompatible with the observed cosmic
isotropy if its present strength were larger than about $10^{-7}$
gauss \cite{Zeldovich}.  Furthermore, it is a general feature of
models predicting magnetic field generation during primordial
phase transitions \cite{models} to forecast random magnetic fields 
at the end of the transition, in domains having a typical size $d
\ll H^{-1}$.  Although magnetohydrodynamical \cite{BEO} and
dissipative effects \cite{Olinto} can
cause the ratio $d/H^{-1}$ to grow during the cosmic expansion it
may still be much smaller than unity at the BBN time.  At that epoch,
the neutrino oscillation length is not much smaller than $H^{-1}$, so
that the neutrino probes a number of field inversions before one spin
precession is complete.

If the magnetic field performs a random walk along each neutrino
trajectory, the average transverse field decreases with the
square root of the length scale.  Therefore, one would expect that
$\gamma=1/2$, whence it appears more natural that $\gamma>3/8$.

In this second generic case the condition \eq{bound} is easiest to
fulfil at early times. Typically, the earliest useful epoch is just
after the QCD phase transition at $T\approx 150\,\rm MeV$. Then,
because $\gamma>3/8$ by assumption, the required value for $\mu B_0$
will be smaller than \eq{X014} by a factor $(T_0/T)^{4\gamma-3/2}$,
which for $\g=1/2$ is an order of magnitude.  Therefore, \eq{X014} is
a conservative requirement in the sense that this value for $\mu B_0$
is certainly sufficient to populate r.h.\ neutrinos before BBN, but a
smaller value may suffice, depending on the exact scaling law of the
effective $B$-field.

In our derivation we have assumed that the effective mixing angle is
small, a condition that we now need to verify. From Eq.~(\ref{X005}) 
we need to demand that 
$2\langle\mu B_{\rm T}\rangle/\langle V_{\rm L}\rangle\alt 1$ or
\be
  \label{X015}
  \mu B_0\alt f_{\rm L}\,\frac{7\pi}{80\,\alpha}\,
  G_{\rm F}^2 T^{3-4\gamma} T_0^{2+4\gamma}~~.
\ee
This condition is most difficult to fulfil at late times, unless
$\gamma>3/4$. 
Therefore, it is enough to
check it at $T=T_0=1\,{\rm MeV}$. At that temperature it amounts to
$\mu B_0\alt 5\times10^{-15}\,\rm eV$. A comparison with
\eqs{X014}{X015} reveals that our assumption of a small mixing angle
has been marginally consistent for $\gamma<3/8$, and safe for
$3/8<\gamma<3/4$.  Assuming that the mixing angle is large amounts to
ignoring refractive effects. This leads to a requirement similar to
Eq.~(\ref{X014}) for r.h.\ neutrinos to reach thermal equilibrium.

The magnetically induced spin-oscillation of neutrinos in the early
Universe has been discussed in several recent papers \cite{Semikozetal}.
While some of them discuss the importance of neutrino refractive effects
at length, this effect does not always matter in their final result. The
difference to our treatment is that we study the effect of correlated
domains with finite sizes while these papers use the limit where the
fields in different points are uncorrelated, $\<B_i(\bx)B_j(\by)\>\sim
\delta_{ij}\delta^{(3)}(\bx-\by)$. In that limit, the magnetic field is
assumed to consist of very small domains of random magnetic field
strength and direction.  Therefore, the main difference between our
discussion and that of Refs.~\cite{Semikozetal} consists of the
assumptions about the magnetic field distribution, and the kinetic
treatment adequate for those assumptions.


\section{Discussion and Summary}
\label{s:disc}

We have studied magnetically induced spin precessions of Dirac
neutrinos in the early Universe. To this end we have derived
expressions for the neutrino dispersion relations in magnetized media
which are valid for field strengths $B$ up to about $m_W^2$. In the
weak-field limit, our results agree with those of D'Olivo, Nieves, and
Pal \cite{DOlivoNP89} apart from an overall sign. In a
charge-symmetric plasma, there is no magnetization contribution to the
neutrino refractive index to lowest order in $m_W^{-2}$, 
contrary to the claim of Semikoz and Valle
\cite{SemikozValle}. Besides a formal derivation, we have shown how to
obtain the magnetic refraction term in a direct and simple physical
fashion, which establishes without ambiguity the absolute sign, and the
relative sign between the electron and positron contributions.

Our analysis indicates that r.h.\ Dirac neutrinos would be thermally
populated by spin oscillations if $\mu B_0\agt 10^{-6}\mu_{\rm
B}\,{\rm gauss}$, where $\mu$ is the assumed neutrino magnetic dipole
moment, $\mu_{\rm B}=e/2m_e$ is the Bohr magneton, and $B_0$ a 
horizon-scale magnetic field at $T_0=1\,\rm MeV$, i.e.\ just before 
the epoch of nucleosynthesis. Depending on the spatial magnetic-field
distribution on smaller scales, i.e.\ with sufficient power in
smaller-scale field modes, even a smaller value of $\mu B_0$
would suffice to thermalize the r.h.\ states. 

In principle, r.h.\ neutrinos could also be populated by direct
spin-flip collisions on charged particles or from annihilation 
processes involving virtual photons  \cite{Morgan}. The 
dipole moment needed to achieve thermal equilibrium for the r.h.
\ states is $\mu\agt0.5\times10^{-10}\mu_{\rm B}$. 
If the neutrino mass is smaller than 1 MeV, as we assume in
the present paper, stellar-evolution bounds on neutrino dipole 
or transition moments are 
$\mu\alt3\times10^{-12}\mu_{\rm B}$ \cite{Raffelt,Dipolebounds},
so that the scattering mechanism cannot be effective in the early
Universe. 

In the particle-physics standard model, neutrinos have no magnetic
dipole moments. However, if neutrinos have a Dirac mass $m$ they
automatically have a dipole moment 
$\mu/\mu_{\rm B}=3.2\times10^{-19}\,m/{\rm eV}$. In other extensions
of the standard model much larger values can be obtained.  
If one of the neutrinos would saturate the stellar-evolution limit,
a primordial field at nucleosynthesis $B_0\approx3\times10^5\,\rm
gauss$ would be enough to populate the r.h.\ degrees of freedom, and
an even smaller field could suffice, depending on its spatial
distribution.

Unfortunately, direct observations of primordial magnetic fields are
still lacking, although it was recently suggested that they may be
detectable by observing their inprint on the cosmic rays \cite{cr} or
on the cosmic microwave radiation \cite{KosLob}.  
However, it may be useful to consider some
recent hypotheses about the genesis and evolution of primordial
magnetic fields.  Many of these propositions are motivated by the
desire to explain the observed galactic and intergalactic magnetic
fields as relics of a primordial cosmological field.  Field strengths
of order $10^{-6}\,\rm Gauss$ are a quite general character of the
interstellar medium. Remarkably, this strength corresponds to an
energy density equal to that of the cosmic microwave background
radiation.  Kronberg \cite{Kronberg} suggests that this feature may be
the result of an early equipartition between magnetic fields and
radiation, a hypothesis that may have found some theoretical support
(e.g.~Ref.~\cite{Baym}).  If this were the case we could expect $B_0
\approx 10^{13}\,\rm Gauss$, a value which is not in contradiction
with primordial nucleosynthesis considerations \cite{GraRub}.  If such
large fields were produced before nucleosynthesis, our result implies
that even a dipole moment as small as about $10^{-19}\mu_{\rm B}$
would be enough to thermalize r.h.\ neutrinos.  Thus, neutrinos with
cosmologically interesting Dirac masses in the eV range would have
sufficiently large dipole moments without further extensions of the
standard model.

It has frequently been argued that the observationally inferred
primordial light-element abundances exclude significant novel
contributions to the cosmic expansion rate of the Universe at the
nucleosynthesis epoch. At the present time, however, new questions
concerning the reliability of the previously inferred abundances of
deuterium and $^3$He have arisen, and the overall consistency of BBN
with all of the observations is not assured
\cite{bbncrisis}. Therefore, at the present time one cannot assume
that the observationally inferred primordial light-element abundances
truly exclude one additional thermally excited neutrino degree of
freedom at the nucleosynthesis epoch.  Therefore, it is not the
ambition of our present study to claim a new exclusion range for 
$\mu B_0$, but rather to illuminate some of the important physical
ingredients needed to understand magnetically induced neutrino spin
oscillations in the early Universe.


\section*{Acknowledgments}

This research was supported, in part, by the European Union contracts
CHRX-CT93-0120 (P.E.\ and G.R.) and SC1*-CT91-0650 (D.G.), by the
NorFA grant No.~96.15.053-O (P.E.\ and D.G.)  and at the
Max-Planck-Institut f\"ur Physik by the Deutsche
Forschungsgemeinschaft grant SFB 375.  We thank J.~Cline,
U.~Danielsson, S.~Davidson, K. Enqvist, G.~Ferretti, K.~Kainulainen, 
D.~Persson, H.R.~Rubinstein and V. Semikoz for helpful
discussions.  G.R.\ acknowledges the hospitality of the Theory
Division at CERN during a visit when part of this work was performed.
We also thank J.~Nieves for confirming our analysis of 
the absolute sign  of the magnetization in \eqs{tadfin}{locbubfin}.

\section*{Note Added}

Before circulating the present paper as an E-print we made a draft
version available to Drs.~J.W.F.~Valle and V.~Semikoz who 
subsequently agreed that
our expression for the magnetically induced refractive index was the
correct one. As a formal response they have now circulated a corrected
version of their derivation \cite{SVerratum}  which explicitly 
confirms our finding.


\appendix

\section{Charged-Fermion Propagator}
\label{a:loop}

In order to calculate the tadpole and bubble diagrams in
Sect.~\ref{s:dr}, we need an explicit expression for the electron
propagator $S(x',x'')$ in the presence of an external magnetic field
and an electronic plasma at non-zero temperature and density.  We shall
use two different methods here: Furry's picture for the local term and
Schwinger's proper-time method for the non-local one. They give the
same result for the local terms, but the Furry-picture result is more
direct to interpret physically. For the non-local terms it would be
considerably more difficult to use the Furry picture.

By the Furry picture we mean that the propagator is constructed
explicitly as a sum over solutions to the Dirac equation in a given
gauge.  For a fermion with mass $m$ and charge $q$ (the electron
having a negative charge $q=-e<0$) the propagator has been constructed
in Refs.~\cite{Persson95,Mak94}. For a magnetic field in the positive
$z$-direction, in the gauge $A_\mu=(0,0,-Bx,0)$, it is given by
\bea
\label{SLL}
    iS(x,x)&=&\sum_{l=0}^\infty
    \int_{-\infty}^{+\infty}\frac{dp_0}{2\pi}
    \int_{-\infty}^{+\infty}\frac{dp_y}{2\pi}
    \int_{-\infty}^{+\infty}\frac{dp_z}{2\pi}
    \left[\frac{i}{p_0^2-E_{l,p_z}^2}-2\pi
      \delta(p_0^2-E_{l,p_z}^2)f_{\rm F}(p_0)\right]\nn
      &&\kern2em\times\biggl\{(p_0\g_0-p_z\g_z+m)
      \Bigl[\sigma_+ I_{l,l}(x,p_y)+\sigma_-I_{l-1,l-1}(x,p_y)\Bigr]
      \nn &&\kern7em
      -i\sqrt{2l|qB|}\,
      \Bigl[\g_+I_{l,l-1}(x,p_y)-\g_-I_{l-1,l}(x,p_y)\Bigr]
      \biggr\}~~,
\eea
where%
\footnote{In our convention three-vectors such as
$\bp=(p_x,p_y,p_z)$ and $\bg=(\g_x,\g_y,\g_z)$ 
are the contravariant components of the corresponding
four-vector and thus have  Lorentz indices $i=1,2,3$ upstairs,
i.e. $p_x=p^1$ etc. We use 
the Minkowski metric ${\rm diag}(+,-,-,-)$ so that $p_i=-p^i$ and
$\g_i=-\g^i$ for $i=1,2,3$.}
$\g_\pm\equiv\half[\g_x\pm \sign(qB)i\g_y]$ and
$\sigma_\pm\equiv\half[1\pm\sign(qB)\sz]$. Note that $\sigma_z$
is understood to mean the Dirac spin 
matrix $\frac{i}{2}[\g_x,\g_y]$.
The Landau levels are labelled by $l$ and their energies are 
$E_{l,p_z}^2=m^2+p_z^2+2|qB|l$.
Further,  $f_{\rm F}(p_0)=f^+_{\rm F}(p_0)\,\Theta(p_0)+
f^-_{\rm F}(p_0)\,\Theta(-p_0)$, where 
$f^\pm_{\rm F}(p_0)=(e^{\pm(p_0-\mu)/T}+1)^{-1}$ are the usual 
occupation numbers for particles and antiparticles  
of a Fermi-Dirac distribution at temperature $T$ and chemical 
potential $\mu$. We have also used
$I_{k,l}(x,p_y)\equiv I_k(x,p_y)I_l(x,p_y)$ with
\be
\label{Indef}
  I_l(x,p_y)=\left( \frac{|qB|}{\pi} \right)^{1/4} 
  \exp \left[-\frac{|qB|}{2}\left(x-\frac{p_y}{qB}\right)^{2}\right]
 \frac{1}{\sqrt{l!}} H_l \left[\sqrt{2|qB|}\left(x-\frac{p_y}
 {qB} \right) \right]~~,
\ee
where $H_l$ is a Hermite polynomial. In the lowest Landau level we
define $I_{-1}=0$ for consistency.  

The $dp_y$ integration can be performed by using the completeness
relation
\be
\label{comp}
   \int_{-\infty}^{+\infty} dp_y\, 
   I_k(x,p_y)I_l(x,p_y)=|qB|\delta_{kl}~~,
\ee
which also removes the $x$-dependence from the r.h.s.\ of \eq{SLL}. 
In the end we are only interested in
the thermal part, coming from the $\delta$-function in \eq{SLL}, so we
drop the vacuum contribution from now on. After 
the $dp_z$ integration has been done using the $\delta$-function 
we find
\bea
\label{SLLfin}
iS(x,x)&=& -\frac{|qB|}{4\pi^2}\int_{-\infty}^{+\infty}dp_0 
    f_{\rm F}(p_0)\,(\g_0p_0+m)
    \nn &&\kern5em\times
    \left(\frac{\Theta(p_0^2-m^2)}{\sqrt{p_0^2-m^2}}\,\sigma_+
    +\sum_{l=1}^\infty 
    \frac{\Theta(p_0^2-m^2-2|qB|l)}{\sqrt{p_0^2-m^2-2|qB|l}}
    \right)~~.  
\eea 
The appearance of the projection operator $\sigma_+$ in \eq{SLLfin}
is related to the fact that there is only one possible spin orientation
in the lowest Landau level ($l=0$).

With this result it is straightforward to calculate expectation 
values like 
\be 
  \< \Psibar(x)\g^i\g_5\Psi(x)\>=
  -\tr[iS(x,x) \g^i\g_5]~~, 
\ee 
where the trace is over $\gamma$-matrices.  
Since $\g_i\g_5$
contains an odd number of $\g$-matrices, only the term in the integrand
in \eq{SLLfin}, which is odd in $p_0$, can contribute.  Evidently it is
zero for a vanishing chemical potential, showing in a more formal way
that the magnetization term of Semikoz and Valle \cite{SemikozValle}
cannot be correct.  

It is often more convenient to label the Landau levels with
an orbital quantum number
$n=0,1,2\ldots$ and a spin
quantum number $\lambda=\pm1$. The 
 energies  are then $E^2_{n,\lambda,p_z}=
m^2+p_z^2+|qB|(2n+1-\lambda)$. For a charged Dirac 
fermion $f$ the net total
number density (particles minus antiparticles) is
\be
\label{S3S4}
    N_{f-\bar{f}}=
    \frac{|qB|}{2\pi^2}\int_0^\infty dp_z\sum_{n=0}^\infty
    \sum_{\lambda=\pm 1}\left[f_{\rm F}^+(E_{n,\lambda,p_z})
    -f_{\rm F}^-(E_{n,\lambda,p_z})\right]~~,
\ee
while the net number density in the lowest Landau level is
\be
   N^0_{f-\bar{f}}=
    \frac{|qB|}{2\pi^2}\int_0^\infty dp_z
    \left[ f_{\rm F}^+(E_{0,1,p_z})-
     f_{\rm F}^-(E_{0,1,p_z})\right]~~.
\ee
These results allow us to relate the local terms of the neutrino
self-energy to the total charge density or to the charge density in the
lowest Landau level, leading to \eqs{tadfin}{locbubfin}.

For the non-local neutrino self-energy term it is convenient
to start from the electron propagator in the Schwinger proper-time 
form, which can be written as \cite{ElmforsPS95,Schwinger51}:
\be
\label{zprop}
        iS(x',x'')=
        \phi(x',x'') \int\frac{d^4p}{(2\pi)^4}\,
         e^{-ip(x'-x'')}i S(p)~~,
\ee
where $\phi(x',x'')$ is a gauge-dependent phase factor.
The gauge-independent and translationally invariant part of 
$S$ is  
\be
\label{thprop}
   i S(p)=iS_{\rm vac}(p)-f_{\rm F}(p_0)
   \Bigl[iS_{\rm vac}(p)-
        iS^*_{\rm vac}(p)\Bigr]~~, 
\ee
where
\bea
\label{Bprop}
        iS_{\rm vac}(p)&=& 
        \int_0^\infty ds\, \frac{e^{iqBs\sigma_z}}{\cos(qBs)}
        \exp\left[is\left(p^2_\parallel-
        \frac{\tan(qBs)}{qBs}\,p^2_\perp-m^2
        +i\ve\right)\right]\nn
        &&\times\left(\gamma p_\parallel-
        \frac{e^{-iqBs \sigma_z}}{\cos(qBs)}\,\gamma p_\perp
        +m\right)\ ,
\eea
where for general four-vectors $a$ and $b$, 
$a\cdot b_\parallel=a_0b_0-(\Bhat\cdot\vek{a})(\Bhat\cdot\vek{b})$ and
$a\cdot b_\perp=\vek{a}\cdot\vek{b}-(\Bhat\cdot\vek{a})(\Bhat\cdot\vek{b})$.
The real combination that occurs in
the thermal part of \eq{thprop} is obtained by extending the
$s$-integral in \eq{Bprop} from $-\infty$ to $+\infty$. In the
integrand of \eq{Bprop} there are poles and essential singularities on
the real $s$-axis. They have to be avoided by taking the integration
contour in the lower half-plane for positive $s$ (see
e.g.\ Ref.~\cite{ElmforsPS94} for a discussion of this contour).
Therefore, to get a real quantity for the thermal part, this contour
has to go in the lower half-plane for negative $s$ as well. 

The $W$ boson propagator has a similar form but with a different
tensor structure. In a closed loop, the gauge-dependent phase factors
$\phi(x,x')$ cancel and the result is explicitly translationally
invariant.  The contribution from thermal $W$ bosons is Boltzmann,
suppressed by a factor $e^{-m_W/T}$ and can be neglected.  Expanding
the $W$ propagator in both the momentum transfer and the $B$ field we
find that the leading $B$-dependent $\cO(m_W^{-4})$-term is local and
that the first non-local $B$-dependent term is $\cO(m_W^{-6})$.
The local term vanishes in a CP symmetric plasma.
Therefore, when calculating the neutrino self-energy to order
$m_W^{-4}$ we may use the zero-field $W$ propagator.

The advantage with the Schwinger proper-time form over the Furry
picture is that the gauge-dependent phase factor disappears
automatically and we do not have to match the wave functions of the
electron propagator (i.e.\ the Landau levels) with the ones of the
$W$ propagator in the zero field limit (i.e.\ plane waves).

With the propagators in \eqs{gbprop}{thprop} it is possible to perform
the loop integral over the three-momenta in \eq{Sigbub} explicitly, but the
result is still fairly complicated. It simplifies considerably in the
linear-field approximation ($B\ll T^2$, $B\ll m^2$), where we have, from
the $W$--$e$-loop: 
\bea
\label{nlweak}
    \Sigma_{\rm bubble}&=&-\frac{g^2}{2m_W^4}
    \int\frac{d^4p}{(2\pi)^4}\,f_{\rm F}(p_0)
    \int_{-\infty}^\infty ds\,e^{is(p^2-m^2)}
    \g^\mu\Bigl[\g p+m+iq\Bhat\cdot\bs s(\g p_\parallel+m)\Bigr]
    \nn&&\kern15em\times
    \Bigl[g_{\mu\nu}(k-p)^2-(k-p)_\mu(k-p)_\nu\Bigr]\g^\nu~~.
    \nn&&
\eea
After adding the $Z$--$\nu$-loop and keeping only the leading
high-temperature piece, we obtain the result in \eq{Signonloc}.
However, \eq{nlweak} is valid also for temperatures lower than the 
electron mass $m$. It contains corrections to \eq{Signonloc}, which 
can be important if $T\simleq m$. 


\end{document}